\documentclass{ws-procs975x65}
\def\draftnote{\relax}
\def\trimmarks{\relax}

\begin{document}
\wstoc{The Quantum Measurement Process:Lessons From An Exactly
Solvable Model}{A. E. Allahverdyan, R. Balian and T. M.
Nieuwenhuizen}

\title{THE QUANTUM MEASUREMENT PROCESS: \\LESSONS FROM AN EXACTLY SOLVABLE MODEL}

\author{Armen E. ALLAHVERDYAN$^{1)}$, Roger BALIAN$^{2)}$ 
AND Theo M. NIEUWENHUIZEN$^{3)}$}
\address
{$^{1)}$Yerevan Physics Institute,
Alikhanian Brothers St. 2, Yerevan 375036, Armenia\\ 
$^{2)}$ Service de Physique Th\'eorique, 
CEA-Saclay, 91191 Gif-sur-Yvette cedex, France\hspace{2cm}{}\\
$^{3)}$Institute for Theoretical Physics, University of Amsterdam\\
Valckenierstraat 65, 1018 XE Amsterdam, The Netherlands}

\begin{abstract} The measurement of a spin-$\frac{1}{2}$ is modeled by coupling it to 
an apparatus, that consists of an Ising magnetic dot coupled to a phonon bath.
Features of quantum measurements are derived from the dynamical solution of the 
measurement, regarded as a process of quantum statistical mechanics. 
Schr\"odinger cat terms involving both the system and the apparatus,
die out very quickly, while the registration is a process taking the apparatus 
from its initially metastable state to one of its stable final states. 
The occurrence of Born probabilities can be inferred at the macroscopic level,
by looking at the pointer alone.  Apparent non-unitary behavior of the measurement process
is explained by the arisal of small many particle correlations,
that characterize relaxation.
\end{abstract}
\keywords
{apparatus, quantum measurement model, spin, registration, collapse, Schr\"odinger cats, decoherence}

\bodymatter

\section{Introduction}

As any new student is taught, the result of a quantum measurement process is coded 
in the collapse postulate and the Born rule. Among theorists,
this formal approach has too often led to the opinions that quantum measurements 
either require a special theory, or that the measurement process 
itself is a piece of physics that need not be worried about --- how different is the
life of an experimentalist, whose task it is to perform the measurement!

It is our aim to discuss features of the quantum measurement process along the lines 
of a rich enough, realistic, but still solvable model. Most of these  points are well
known in literature, while a few other ones are put forward by the solution of the
problem. In particular, it is shown that the collapse is explained by a dynamical 
approach relying only on the Schr\"odinger equation and on statistical properties
issued from the large size of the apparatus.

The most standard theory of measurements is the von Neumann-Wigner theory,
where the apparatus is described in terms of pure states.~\cite{WheelerZurek,Schlosshauer}
 We consider that
a sensible approach to the problem should rely on quantum statistical physics. 
Following other works in the literature
~\cite{Hepp,Cini,Gaveau, Schulman,Haake,Venugopalan,Mozyrsky, ABNmeasPRA,Braun,
ABNmeasEPL,Sewell,Merlin},
we consider an explicit model for the measurement apparatus. 

We shall first outline
the model, and then highlight various relevant aspects of the process.
The measurement that we describe is ideal (or non-demolishing, or of the first kind)
in the sense that the sole perturbation
brought to the system by the dynamical process is von Neumann's reduction of the state.

We keep aside all derivations and technical details, one or the other 
can be found in our earlier papers on this model.
\cite{ABNmeasEPL,ABNmeasNaples,ABNmeasVaxjo04,ABNmeasPrague,ABNmeasVaxjo05}.

\section{The model and its solution}

For the tested system $\mathrm{S}$, we take the simplest quantum system, that is, a spin
$\frac{1}{2}$. Its initial state is represented by the $2\times2$ density matrix $\hat r(0)$.
 The observable to be measured is its third Pauli
matrix $\hat{s}_{z}$ with eigenvalues $s_{i}$ equal to $\pm1$. 

The interaction of S with the apparatus A should trigger A from some initial state $\hat{\cal R}(0)$
into either one of two possible final states  $\hat{\cal R}_\Uparrow$ or $\hat{\cal R}_\Downarrow$,
associated with $s_\uparrow=+1$ or $s_\downarrow=-1$, respectively.
(Here $\Uparrow$ and $\Downarrow$ refer to A, while $\uparrow$ and $\downarrow$ refer to S.)
These states $\hat{\cal R}_\Uparrow$ and $\hat{\cal R}_\Downarrow$ should have no overlap,
and should be distinguishable at our scale by observation of some pointer variable.
Registration in either one should be permanent and insensitive to weak external perturbations.
Symmetry between $\hat{\cal R}_\Uparrow$ and $\hat{\cal R}_\Downarrow$ should prevent any bias.

In order to satisfy these conditions, we take for A a system that can undergo a phase transition 
with broken invariance. Attempting to conciliate mathematical tractability and realistic features, 
we choose A as the simplest quantum object that displays two phases, namely a sufficiently large 
Ising system. Though finite, the number of degrees of freedom will be
sufficiently large so that the relaxation towards one of the equilibrium
states $\hat{\mathcal{R}}_{\Uparrow}$, 
$\hat{\mathcal{R}}_{\Downarrow}$ is irreversible and that the order parameter
 has weak fluctuations in each possible final state.
The transition of $\mathrm{A}$ from a metastable state to one of its 
stable, macroscopic states
eliminates the infamous problem of observers (``Wigner's friend'', ``Mind-body problem''),
see e.g. ~\cite{WheelerZurek}, from the quantum measurements process: 
After the measurement, the pointer variable 
will have a stable value, and can be read off at any moment, or just not, 
without causing a back-reaction on the already finished measurement process.

Our apparatus $\mathrm{A}=\mathrm{M}+\mathrm{B}$ 
simulates a \textit{magnetic dot}: the magnetic degrees of freedom $\mathrm{M}$
consist of $N\gg1$ spins with Pauli operators $\hat{\sigma}_{a}^{\left(
n\right)  }$, ($a=x$, $y$, $z$), while the non-magnetic degrees of freedom
such as phonons behave as a thermal bath $\mathrm{B}$. Anisotropic
interactions between these spins can generate Ising ferromagnetism below the
Curie temperature $T_{\mathrm{C}}$. As pointer variable we take the
order parameter, which is the magnetization in the $z$-direction, represented
(within normalization)\ by the quantum observable
\begin{equation}
\hat{m}=\frac{1}{N}\sum
_{n=1}^{N}\hat{\sigma}_{z}^{\left(  n\right)  }\text{ .}\label{hatm=}%
\end{equation}
The initial state $\mathcal{\hat{R}}\left(  0\right)  $\ of $\mathrm{A}$\ is
the metastable paramagnetic state $(\langle\hat m\rangle=0)$, prepared by first thermalizing
$\mathrm{A}=\mathrm{M}+\mathrm{B}$ at a temperature $T_{0}$ above
$T_{\mathrm{C}}$, then suddenly cooling $\mathrm{B}$ into equilibrium at the
temperature $T$\ below $T_{\mathrm{C}}$. We expect the final state of the process
to involve for $\mathrm{A}$ the two stable ferromagnetic states
$\mathcal{\hat{R}}_\Uparrow$ or $\mathcal{\hat{R}}_\Downarrow$, with broken invariance. 
The equilibrium
temperature $T$\ will be imposed to $\mathrm{M}$\ by the phonon bath through
weak coupling between the magnetic and non-magnetic degrees of freedom. Within
fluctuations small as $1/\sqrt{N}$, the order parameter Eq. \ref{hatm=} vanishes
in $\mathcal{\hat{R}}\left(  0\right)  $\ and takes two opposite values\ in
the states $\mathcal{\hat{R}}_{\Uparrow}$ and $\mathcal{\hat{R}}_{\Downarrow}$, 
$\left\langle \hat{m}\right\rangle _{i}$ equal to
$+m_{\mathrm{F}}$ for $i=\Uparrow$ and to $-m_{\mathrm{F}}$ for $i=\Downarrow$. 
As in real magnetic registration devices, information will be stored by 
$\mathrm{A}$\ in the form of the sign of the final magnetization.

\subsection{The Hamiltonian}

The full Hamiltonian can be decomposed into terms associated with the system,
with the apparatus and with their coupling:%
\begin{equation}
\hat{H}=\hat{H}_{\mathrm{S}}%
+\hat{H}_{\mathrm{A}}+\hat{H}_{\mathrm{SA}}\text{ .}\label{ham}%
\end{equation}
In an ideal measurement the observable $\hat{s}$ to be measured should not
change during the process, so that it should commute with $\hat{H}$. 
The simplest self-Hamiltonian that ensures this
property is the trivial one: $\hat{H}_{\mathrm{S}}=0$.
The coupling between the tested system and the apparatus is a spin-spin coupling,%
\begin{equation}
\hat{H}_{\mathrm{SA}}=-g\hat
{s}_{z}\sum_{n=1}^{N}\hat{\sigma}_{z}^{\left(  n\right)  }=-Ng\hat{s}_{z}%
\hat{m}\text{ .}\label{hamham}%
\end{equation}
Before the measurement and after it, $g$ will be  equal to zero.

The apparatus $\mathrm{A}$ consists, as indicated above, of a magnet
$\mathrm{M}$ and a phonon bath $\mathrm{B}$, and its Hamiltonian can be
decomposed into%
\begin{equation}
\hat{H}_{\mathrm{A}}=\hat
{H}_{\mathrm{M}}+\hat{H}_{\mathrm{B}}+\hat{H}_{\mathrm{M}\mathrm{B}}\text{
.}\label{HA}%
\end{equation}
The magnetic part is chosen as the long-range Ising interaction,
\begin{equation} 
\hat{H}_{\mathrm{M}}=-\frac{1}{2}JN\hat{m}^{2}\text{ .}
\label{HM=}%
\end{equation}
The magnet-bath interaction, needed to drive the apparatus to equilibrium, is taken
as a standard spin-boson Hamiltonian%
\begin{equation}
\hat{H}_{\mathrm{M}\mathrm{B}%
}=\sqrt{\gamma}\sum_{n=1}^{N}\left(  \hat{\sigma}_{x}^{\left(  n\right)  }%
\hat{B}_{x}^{\left(  n\right)  }+\hat{\sigma}_{y}^{\left(  n\right)  }\hat
{B}_{y}^{\left(  n\right)  }+\hat{\sigma}_{z}^{\left(  n\right)  }\hat{B}%
_{z}^{\left(  n\right)  }\right)  
\text{ ,}\label{ham4}%
\end{equation}
which couples each component $a=x$, $y$, $z$ of each spin $\mathbf{\hat
{\sigma}}^{\left(  n\right)  }$ with some hermitean linear combination
$\hat{B}_{a}^{\left(  n\right)  }$ of phonon operators. The dimensionless
constant $\gamma\ll1$ characterizes the strength of the thermal coupling
between $\mathrm{M}$ and $\mathrm{B}$, which is weak.

The bath Hamiltonian $\hat{H}_{\mathrm{B}}$ is a large set of harmonic oscillators.
It will be involved in our problem only through
its \textit{autocorrelation function} in the equilibrium state 
at temperature $T=1/\beta$,
defined by%
\begin{eqnarray} &&
\frac{1}{Z_{\rm B}}\operatorname{tr}_{\mathrm{B}}
\left[\,e^{-\beta\hat H_{\rm B}}
\hat{B}_{a}^{\left(  n\right)  }\left(  t\right)  \hat{B}_{b}^{\left(
p\right)  }\left(  t^{\prime}\right)  \right]  =\delta_{n,p}\delta
_{a,b}\,K\left(  t-t^{\prime}\right)  \text{ ,}\label{tK-s=}\\
&&\hat{B}_{a}^{\left(  n\right)
}\left(  t\right)   \equiv
e^{i\hat{H}_{\mathrm{B}}t/\hbar}
\hat{B}_{a}^{\left(  n\right)  }e^{-i\hat{H}_{\mathrm{B}}t/\hbar},
\end{eqnarray}
in terms of the evolution operator of $\mathrm{B}$ alone. 
 We choose for our model as
Fourier transform%
\begin{equation}
\tilde{K}\left(  \omega\right)
=\int_{-\infty}^{+\infty}\mathrm{d}t\ e^{-i\omega t}K\left(  t\right)
\label{TF}%
\end{equation}
of $K\left(  t\right)  $ the simplest expression having the required
properties, namely%
\begin{equation}
\tilde{K}\left(
\omega\right)  =\frac{\hbar^{2}}{4}\frac{\omega e^{-\left\vert \omega
\right\vert /\Gamma}}{e^{\beta\hbar\omega}-1}\text{ ,}\label{K^tilde}%
\end{equation}
known as a quasi-Ohmic spectrum.
The temperature dependence accounts for the quantum bosonic nature of the
phonons. The Debye cutoff $\Gamma$ characterizes the largest frequencies of
the bath, and is assumed to be larger than all other frequencies entering our
problem. 

\subsection{Disappearance of Schr\"odinger cats}

\subsubsection{Dephasing}

The full density operator $\hat{\cal D}(t)$ of S+A is initially factorized as
\begin{equation} \hat{\cal D}(0)=\hat{ r}(0)\otimes \hat{\cal R}(0),\qquad \end{equation}
with S described by \begin{equation} 
\label{rhat(0)}
\hat r(0)=\begin{pmatrix}
r_{\uparrow\uparrow}(0) & r_{\uparrow\downarrow}(0)    \\
r_{\downarrow\uparrow}(0) & r_{\downarrow\downarrow}(0)
\end{pmatrix}, 
\end{equation}
and A by $ \hat{\cal R}(0)$. $\hat{\cal D}(t)$ 
evolves according to the Liouville-von Neumann equation. We have exactly
solved this equation, for values of the parameters of the model satisfying

\begin{equation} N\gg1, \qquad\hbar\Gamma\gg T\gg \gamma J\gg 
\frac{J}{N}\left(\frac{g}{\hbar\Gamma}\right)^2,
\qquad \hbar\Gamma\gg J>g.\end{equation}

Over the very brief time-scale
\begin{equation} \tau_{\rm red}=\frac{1}{\sqrt{2N}}\,\frac{\hbar}{g},\end{equation}
the off-diagonal blocks $\hat\Pi_\uparrow \hat{\cal D}(t)\hat\Pi_\downarrow$ and
$\hat\Pi_\downarrow \hat{\cal D}(t)\hat\Pi_\uparrow$, where 
$\hat\Pi_\uparrow=|\!\!\uparrow\rangle\langle\uparrow\!\!| $ 
and $\hat\Pi_\downarrow=|\!\!\downarrow\rangle\langle\downarrow\!\!|$ 
denote the projection operators on the subspaces 
$s_\uparrow=+1$ and $s_\downarrow=-1$, respectively, decay to zero.
This process takes place on a timescale so short that energy exchange with 
the apparatus is negligible, so the bath does not play a role in it.
Indeed, what occurs is a dephasing process 
caused by the interaction of the tested spin with the $N$ spins of the apparatus. 
Just as in a spin-echo setup, due to the phase coherence, the initial state could in 
principle be retrieved,
but this should not occur if we wish the process to be used as a quantum
measurement. Recurrences might appear on an extremely long timescale
$\tau_{\rm recur}$ and, in principle, some mechanism is needed to exclude them.
Thus, the so-called ``Schr\"odinger cat'' terms, superposing up and down projections, 
that have so much troubled the understanding of the quantum measurement process, 
just die out in the initial stage of the process.

\subsubsection{Decoherence}

The definitive nature of this process (irreversibility) is ensured by the large number of degrees of 
freedom of the apparatus, which prevents recurrences to occur over any reasonable timescale.
In the model this is ensured by the bath. It starts to play a role at a time scale
\begin{equation}
\tau_{\mathrm{irrev}}^{\mathrm{B}}=
\left[  \frac{2\pi\hbar^{2}}{N\gamma g^{2}\Gamma^{2}}\right]^{1/4}. \
\label{tauBirr2}%
\end{equation}

We choose the parameters of the model, in particular, the coupling $\gamma$
between M and B, in such a way that
\begin{equation}
\tau_{\mathrm{red}}\ll\tau_{\mathrm{irrev}}\ll\tau_{\mathrm{recur}},
\end{equation}
that is,
\begin{equation}
N\gg\frac{\gamma\hbar^{2}\Gamma^{2}}{8\pi g^{2}}\gg\frac{4}{N\pi^{4}}.
\label{condgamma}%
\end{equation}
These conditions are easy to satisfy for large $N$.

\subsection{Registration of the measurement}
 
Only the diagonal blocks $\hat\Pi_\uparrow \hat{\cal D}(t)\hat\Pi_\uparrow$ 
and $\hat\Pi_\downarrow \hat{\cal D}(t)\hat\Pi_\downarrow$  are then left,
and it takes a much longer time,

\begin{equation}
\tau_{\rm reg}=\frac{\hbar}{\gamma(J-T)}\,\ln\frac{3m_{\rm F}(J-T)}{g},
\end{equation}
for the apparatus to register the measurement through the evolution of these
diagonal blocks. We find that the final state of the
compound system S+A is represented, at a time $t_{\rm f}>\tau_{\rm reg}$, by
von Neumann's reduced density operator
\begin{equation} 
{\hat{\mathcal{D}}}\left(
{t_{\mathrm{f}}}\right)  =
r_{\uparrow\uparrow}(0)|\!\uparrow\rangle\langle\,\uparrow\!|  
\otimes\hat{\mathcal{R}}_{\Uparrow}+
r_{\downarrow\downarrow}(0)|\!\downarrow\rangle\langle\downarrow\!|  
\otimes\hat{\mathcal{R}}_{\Downarrow}\,
\text{,}\label{Dtf=}
\end{equation}
where $\hat{\mathcal{R}}_{\Uparrow}$ and $\hat{\mathcal{R}}_{\Downarrow}$
represent the two stable ferromagnetic states of the apparatus.
The success of the measurement also requires the lifetime of the 
initial, unstable paramagnetic state $\hat{\mathcal{R}}(0)$ of A to be 
longer than the duration of the measurement, a condition which 
is satisfied in the model.\cite{ABNmeasVaxjo05}
In earlier work,
\cite{ABNmeasEPL,ABNmeasNaples,ABNmeasVaxjo04,ABNmeasPrague}
 we considered a quartic interaction instead of the quadratic
interaction Eq. \ref{HM=}~\cite{merv}; since the transition is then of first order,
the lifetime of the paramagnetic state is even much longer.

The magnetization $m$ is a macroscopic variable, which for large $N$ behaves 
continuously with small statistical fluctuations $\sim1/\sqrt{N}$.
The quantity 
\begin{equation} \label{p(m)=} p(m;t_{\rm f})={\rm Tr}\,\hat{\cal D}(t_{\rm f})\, \delta(m-\hat m)
=p_\Uparrow\,\delta(m-m_{\rm F})+p_\Downarrow\,\delta(m+m_{\rm F}),
\end{equation}
derived from Eq. \ref{Dtf=}, can only be interpreted as the probability
distribution for $m$ at the time $t_{\rm f}$.
In repeated measurements, the prefactors
\begin{equation} \label{p=r} p_\Uparrow=r_{\uparrow\uparrow}(0), \qquad 
p_\Downarrow=r_{\downarrow\downarrow}(0),
\end{equation}
are the frequencies with which 
we shall observe magnetizations $+m_{\rm F}$ and $-m_{\rm F}$, respectively.
Because of this connection, the measurement is called {\it ideal}.
Born's rule is therefore derived through the dynamical 
analysis of the measurement process instead of being posed as a postulate.
Indeed, Born probabilities in the frequency interpretation are obtained
straightforwardly from Eq. \ref{p(m)=}: since this equation refers to the pointer
variable alone, which is a macroscopic variable, only the above
identification can be given for its meaning. After this has been set, the
same meaning must hold for Eq. \ref{Dtf=} and for the post-measurement state of
the tested system,
\begin{equation} \label{rhat(tf)}
\hat r(t_{\rm f})=r_{\uparrow\uparrow}(0)|\!\uparrow\rangle\langle\,\uparrow\!|  
+ r_{\downarrow\downarrow}(0)|\!\downarrow\rangle\langle\downarrow\!|
=\begin{pmatrix}
r_{\uparrow\uparrow}(0) & 0    \\
0 & r_{\downarrow\downarrow}(0)
\end{pmatrix}.
\end{equation}
This expression thus represents a distribution over an ensemble of measured
spins in the frequency interpretation, 
a simpler result than anticipated, see e.g. ~\cite{Cox,deFinetti,Caves}
 It seems incompatible with, e.g.,
the Copenhagen interpretation or a many world interpretation.

However, the expression Eq. \ref{Dtf=} for the final state of S+A contains additional
information: it exhibits a full correlation between the apparatus and the measured spin,
and involves no Schr\"odinger cat term.
The disappearance of such terms has resulted from the exact solution of the model,
and did not rely on any ``collapse postulate'' or ``projection''.
 
\section{Microscopic versus macroscopic aspects of quantum measurement}

Several important features of quantum measurements are put 
forward by the solution of the above model.

The apparatus $\mathrm{A}$ should \textit{register} the
result in a robust and permanent way, so that it can be read off by
any observer. Such a registration, which is often overlooked in the literature
on measurements, takes place during the second stage of the process as indicated
above. It
is needed for practical reasons especially since $\mathrm{S}$
is a microscopic object. Moreover, its very existence allows us to disregard
the observers in quantum measurements. Once the measurement has been
registered, the result becomes objective. 
The literature which attributes a r\^ole to the mind who observes S is 
therefore irrelevant. 

Registration also requires an \textit{amplification} within the
apparatus of a signal produced by interaction with the microscopic system
$\mathrm{S}$. For instance, in a bubble chamber, the apparatus in its initial
state involves a liquid, overheated in a metastable phase. In spite of the
weakness of the interaction between the particle to be detected and this
liquid, amplification and registration of its track can be achieved owing to
local transition towards the stable gaseous phase. This stage of the
measurement process thus requires an irreversible phenomenon. It is governed
by the kinetics of bubble formation under the influence of the particle and
implies a dumping of free energy in the surrounding liquid, the dynamics of
which governs the size of the bubble. Similar remarks hold for photographic
plates, photomultipliers or other types of detectors.
In our model the amplification is ensured by the interaction between the
magnet M and the phonon bath B, which allows the energy exchange  
and the entropy increase needed to bring M from the state 
$\hat{\cal R}(0)$ to $\hat{\cal R}_\Uparrow$ or $\hat{\cal R}_\Downarrow$.

A measurement process thus looks like a {\it relaxation process},
but with several complications. On the one hand, the final state of
$\mathrm{A}$ is not unique, and the dynamical process can have several
possible outcomes for A. The apparatus is therefore comparable to a material which
 has a broken invariance, and can relax towards one
equilibrium phase or another, starting from a single metastable phase. 
This is why we chose a model involving such a phase transition. 

All these features, registration, amplification, existence of several possible 
outcomes, thus require the apparatus to be a macroscopic object, whereas 
S is microscopic.

 \section{Measurement, a process of quantum statistical mechanics}

The large size of A cannot be dealt with by any other means than statistical mechanics.
Of course, we must treat S+A as a compound {\it quantum} system.
The use of statistical mechanics compels us to regard our description of a measurement 
as relevant to a statistical ensemble of processes, not to an individual process.
In particular, our equation Eq. \ref{Dtf=} for the final state 
${\hat{\mathcal{D}}}\left({t_{\mathrm{f}}}\right)$ defines expectation values, correlations,
possibly with small fluctuations, and this feature is imposed both by quantum mechanics 
and by statistical mechanics for A. This is why we have described the above solution in terms 
of density operators, not of pure states. In particular, we saw that the initial,
metastable state of A was prepared 
by controlling a macroscopic parameter, the temperature of B. 
It is thus represented in the quantum formalism by a
\textit{mixed state}, coded in its density operator $\hat{\mathcal{R}}\left(
0\right)  $. (It is impossible in an actual experiment to make a complete
quantum preparation of a large object, and the assumption that
$\mathrm{A}$ might lie initially in a pure state and
end up in one among some pure states corresponds to a very unrealistic 
thought measurement --- nevertheless this assumption is frequent in measurement theory, see
e.g.~\cite{WheelerZurek,Schlosshauer}.) 
Likewise, each of the final states $\hat{\cal R}_i$, 
characterized by the value of the pointer variable that will be
observed, must again be described by means of a density operator
$\hat{\mathcal{R}}_{i}$, and not by means of pure states as in 
the von Neumann-Wigner approach, too often followed in the literature.
Indeed, the number of state vectors associated with a sharp value of the
\textit{macroscopic} pointer variable is huge for any actual
measurement: As always for large systems, we must allow for small
fluctuations, negligible in relative value,\ around the mean value
$\pm m_{\rm F}$ of $\hat m$.

However, the evolution of $\mathrm{A}$ towards one among its possible final states
$\hat{\mathcal{R}}_{i}$ is 
{\it triggered by interaction with} $\mathrm{S}$, in a way depending on the initial
microscopic state of $\mathrm{S}$ and, for an ideal measurement, the outcome
 should be correlated to the final microscopic state of $\mathrm{S}$,
a property exhibited by the form Eq. \ref{Dtf=} of the final state
$\hat{\cal D}(t_{\rm f})$ of S+A.
Thus, contrary to theories of standard relaxation processes, 
the theory of a measurement process requires a simultaneous
control of microscopic and macroscopic variables. 
In order to solve the coupled equations of motion for 
$\mathrm{A}$ and $\mathrm{S}$ which involve reduction and registration,
we made use of a kind of coarse graining, which is adequate for A, 
becoming exact in the limit of a large $\mathrm{A}$, but 
we had to treat the small system $\mathrm{S}$ exactly. 

Moreover, the final state of S+A keeps {\it memory} of the initial state of S,
at least partly. The very essence of a measurement lies in this feature,
whereas memory effects are rarely considered in standard relaxation processes.

\section{Irreversibility}

A quantum measurement is irreversible for two reasons. On the one hand,
the {\it loss of the off-diagonal blocks}, exhibited in the expression Eq. \ref{Dtf=}
of the final density of S+A, requires an irreversibility of the process.
Even if the initial state were pure, a final state involving only diagonal projections 
would be a statistical mixture -- this irreversibility is associated with the loss
of specifically quantum correlations between $\hat s_x$ or $\hat s_y$ and A, 
embedded in the off-diagonal blocks. On the other hand, the registration by the apparatus
requires an irreversible {\it relaxation} from the metastable paramagnetic state 
$\hat {\cal R}(0)$ towards the stable ferromagnetic states $\hat{\cal R}_\Uparrow$
or $\hat{\cal R}_\Downarrow$, as in the ordinary dynamics of a phase transition.

The irreversibility of the transformation leading from ${\hat{\mathcal{D}}%
}\left(  0\right)  $ to ${\hat{\mathcal{D}}}\left(  {t_{\mathrm{f}}}\right)  $
is measured by the entropy balance. The
von Neumann entropy of the initial state is split into
contributions from $\mathrm{S}$ and $\mathrm{A}$, respectively, as%
\begin{equation}
S\left[  {\hat{\mathcal{D}}%
}\left(  0\right)  \right]  =-\mathrm{\operatorname{tr}}{\hat{\mathcal{D}}%
}\left(  0\right)  \,\ln{\hat{\mathcal{D}}}\left(  0\right)  =S_{\mathrm{S}%
}\left[  \hat{r}\left(  0\right)  \right]  +S_{\mathrm{A}}\left[
\hat{\mathcal{R}}\left(  0\right)  \right]  \text{ ,}\label{entropie0}%
\end{equation}
whereas that of the final state Eq. \ref{Dtf=} is%
\begin{equation}
S\left[  {\hat
{\mathcal{D}}}\left(  {t_{\mathrm{f}}}\right)  \right]  =S_{\mathrm{S}}\left[
\sum_{i}\hat{\Pi}_{i}\hat{r}\left(  0\right)  \hat{\Pi}_{i}\right]  +\sum
_{i}p_{i}S_{\mathrm{A}}\left[  \hat{\mathcal{R}}_{i}\right]  \text{
.}\label{entropiefin}%
\end{equation}
The increase of entropy from Eq. \ref{entropie0} to Eq. \ref{entropiefin}
clearly exhibits the two above-mentioned reasons. On the one hand, when the initial
density operator $\hat{r}\left(  0\right)  $ involves off-diagonal blocks
$\hat{\Pi}_{i}\hat{r}\left(  0\right)  \hat{\Pi}_{j}$ ($i\neq j$), their
truncation raises the entropy. On the other hand, a robust registration
requires that the possible final states $\hat{\mathcal{R}}_{i}$ of
$\mathrm{A}$ are more stable than the initial state $\hat{\mathcal{R}}\left(
0\right)  $, so that also their entropy is larger. The latter effect dominates
because the apparatus is large.

An apparatus is a device which allows us to {\it gain some information} on the state
of $\mathrm{S}$ by reading on A the outcome $+m_{\rm F}$ or $-m_{\rm F}$. 
The price we have to pay, for
being thus able to determine the matrix elements $r_{\uparrow\uparrow}(0)$
and $r_{\downarrow\downarrow}(0)$ referring to $s_z=+1$ and $-1$,
are a complete {\it loss of information} about the off-diagonal elements 
$r_{\uparrow\downarrow}(0)$ and $r_{\downarrow\uparrow}(0)$
of the initial state of
$\mathrm{S}$, and a rise in the thermodynamic entropy of the apparatus.

The solution of our model shows that the so-called ``measurement problem'',
to wit, the fact that the final state Eq. \ref{Dtf=} does not seem to be related
unitarily to the initial state, has the same nature as the celebrated 
``paradox of irreversibility'' in statistical mechanics, \cite{BalianBook}
with additional quantum features. Here too, it is the large size of the apparatus
which produces destructive interferences, thus generating {\it inaccessible 
recurrence times}; such times behave as exponentials of the level density,
which itself is an exponential of the number of degrees of freedom.
As in the solution of the irreversibility paradox,\cite{MayerMayer} we witness here a 
{\it cascade} of correlations involving more and more spins of M, 
towards which the initial order embedded in the off-diagonal elements
of $\hat r(0)$ escapes without any possibility of retrieval.
Mathematically speaking, such correlations should be included in the
final state, but the expression Eq. \ref{Dtf=} is physically exact in the sense
that many-spin correlations cannot be detected and have no observable implication.

\section{Meaning of von Neumann's reduction and of Born's rule}

The solution of our model shows that the disappearance of the off-diagonal elements 
of $\hat{\cal D}(t)$ is a {\it real dynamical phenomenon}, involving 
an irreversible process. Indeed, we found the collapsed final state Eq. \ref{Dtf=}
by merely working out the unitary Liouville-von Neumann equation,
without any other extra ingredient than statistics.

It should be stressed that this disappearance concerns the overall system S+A,
and not S or A separately:
von Neumann's reduction is a property of the {\it compound system} S+A,
which arises for an ideal measurement.  
 In fact, if we take the trace over the system S,
which means that we are no longer interested in S after the time $t_{\rm f}$
but only in the indications of the apparatus as in Eq. \ref{p(m)=}, 
off-diagonal blocks of  $\hat{\cal D}(t)$, even if they were present, would drop out.
Likewise, tracing over A would yield as marginal density matrix 
$\hat r(t)$ for S simply the diagonal elements of $\hat r(0)$,
even if Eq. \ref{Dtf=} had included  off-diagonal blocks.

The elimination of the off-diagonal blocks,
not only for the marginal density matrix $\hat r$ of S,
but also from the overall density matrix $\hat{\cal D}$ of S+A,
contrasts with what happens in usual decoherence processes (for a review on decoherence
processes see \cite{Schlosshauer,Guilini}).
There, a weak interaction of a system with its environment,
which behaves as a thermal bath, destroys off-diagonal blocks in the density matrix
of the system, but the back reaction of this system on its environment is usually not
considered. 
Here, the reduction on the timescale (13) is the result of the interaction (3)
of S with the pointer M, without intervention of the bath B, and the whole
properties of S+A after this process are of interest, including the correlations
$\langle\hat s_x\hat m^k\rangle$ and $\langle\hat s_y\hat m^k\rangle$ for $k\ge 1$, 
which, after increasing initially, vanish on the reduction timescale. 
Because the latter timescale is not related to the bath,
it is misleading to regard reduction as a decoherence process.

Born's rule also involves both S and A.
As exhibited in the expression Eq. \ref{Dtf=}
of the final state, it means that the outcome of the measurement, namely
$+m_{\rm F}$ associated with $\hat{\cal R}_\Uparrow$, or 
$-m_{\rm F}$ associated with $\hat{\cal R}_\Downarrow$, is {\it fully correlated}
in the ideal measurement that we consider with the final state 
$|\!\!\uparrow\rangle$ or $|\!\!\downarrow\rangle$ of S.
We noted moreover that the frequency $p_\Uparrow$ of the occurrence of
$+m_{\rm F}$ in repeated measurements, exhibited in Eq. \ref{p(m)=},
is equal to the number $r_{\uparrow\uparrow}(0)={\rm Tr}\hat\Pi_\uparrow\hat r(0)$.
For an ideal measurement it is also equal to the element
$r_{\uparrow\uparrow}(t_{\rm f})$ of the marginal density operator of S.
\, From the macroscopic character of $p_\Uparrow$ (i.e. being related to
the possible values of the pointer), we can thus infer that 
$r_{\uparrow\uparrow}(0)=r_{\uparrow\uparrow}(t_{\rm f})$
can be interpreted as a probability for the microscopic system S 
to lie in the $+z$-direction in the final state.
\, From $p_\Uparrow$  and $p_\Downarrow$ we also get through   
$r_{\uparrow\uparrow}(0)$ and $r_{\downarrow\downarrow}(0)$, related by Eq. \ref{p=r}, 
partial probabilistic information about the state which describes initially the considered 
statistical ensemble.

When a statistical ensemble of quantum systems is
described by a pure state, any sub-ensemble is described by the same pure
state. On the contrary, for a mixture, we can split the statistical ensemble
into sub-ensembles described by different states, pure or not, provided we
have collected the information needed to distinguish which sub-ensemble
 each system belongs to. This is precisely what happens in the final state
Eq. \ref{Dtf=}\ of our ideal measurement. The outcome registered by A
 provides us with\ the criterion required to sort the successive
experiments into subsets labeled by $\uparrow$ or $\downarrow$. 
The ensemble for which
$\mathrm{S}+\mathrm{A}$ has the density operator Eq. \ref{Dtf=}, is thus split
into sub-ensembles, in each of which $\mathrm{S}$ and $\mathrm{A}$\ are
decorrelated and $\mathrm{S}$\ lies in the eigenstate 
$|\!\uparrow\rangle$ or $|\!\downarrow\rangle$ of $\hat s_z$.
It is because the
final state of $\mathrm{S}+\mathrm{A}$ is a mixture,
owing to the physical reduction,
and because $\mathrm{A}%
$\ relaxes towards either one of the states $\mathcal{R}_{i}$ without
Schr\"odinger cats, that we can use
an ideal measurement process as a \textit{preparation} of $\mathrm{S}$,
initially in $\hat r(0)$, into a new state controlled by the filtering 
of the outcome of A.~\cite{bv}

\section{Relation to the pre-measurement}

In the {\it pre-measurement} stage, information is transfered from the microscopic
system to the macroscopic measuring apparatus, generally followed by a process of 
amplification in the apparatus.~\cite{Peres} 
In a Stern-Gerlach experiment the pre-measurement stage is related to 
the region of space where the magnetic field is inhomogeneous and the individual particles
``decide'' either to the upper or to the lower beam, see e.g. \cite{deMuynck} for a discussion.
 After this, the detection by a remote
detector is somewhat trivial if only the position and not the spin is measured.
Likewise, in our model, one may look for a stage where it is determined 
that the tested particle ends up with its spin either up or down.
Clearly, this should happen in the dephasing process, which 
takes place on the shortest timescale relevant to the measurement,
the reduction time $\tau_{\rm red}$.
Here the single tested spin interacts with the $N\gg1$ spins
of the apparatus, while the bath is still ineffective.

To produce evidence for this connection with pre-measurement,
let us decide to stop the process after the
reduction stage, but before the bath sets in, so on a timescale
$\tau_{\mathrm{red}}\ll\tau\ll\tau_{\mathrm{irrev}}$.
Clearly, the measurement has not been performed since no registration
has taken place. The apparatus is still in the paramagnetic state, 
but small multiparticle correlations have been developed with the tested spin.
Removing the apparatus amounts to trace it out, thus neglecting these
anyhow tiny correlations. However, for the tested spin we end up with the mixture
Eq. \ref{rhat(tf)}, and no recurrence will subsequently occur if the coupling with the
apparatus is removed.
Comparing with the initial state Eq. \ref{rhat(0)}, it is seen that the Schr\"odinger 
cat terms have indeed already disappeared in this stage.
This implies a physics different from the initial state.
If, after this stop of the measurement, the spin is measured again along
the $z$-axis with another apparatus, this will still end up as it was described above.
But if in the second measurement the spin is measured along the $x$-axis,
the outcome for $\langle\hat s_x\rangle$
will not be  $r_{\uparrow\downarrow}(0)+r_{\downarrow\uparrow}(0)$,
but just zero, and likewise for  $\langle\hat s_y\rangle$.
More precisely, we shall find for repeated measurements of $s_x$  or $s_y$ 
the values $+1$ or $-1$ with the same probability, in contrast with
measurements on the initial state Eq. \ref{rhat(0)}.

\section{Statistical interpretation of quantum mechanics}

The very concept of a physical quantity is related in quantum mechanics to 
the possibility of its measurement.
Our model, aimed at understanding measurements as a quantum dynamical process,
 has compelled us to work in the framework of quantum statistical
mechanics, using density operators rather than pure states. 
The process that we described thus refers to a {\it statistical ensemble}
of measurements on a statistical ensemble of systems, 
not to a single measurement experiment.

It is often argued (``ignorance interpretation'')
that a statistical mixture $\mathcal{\hat{D}}$,\ characterizing
only our knowledge of the system,\ should be interpreted as a collection
 of\ \textquotedblleft underlying pure states\textquotedblright%
\ $\left\vert \varphi_{k}\right\rangle $\ which would have more
\textquotedblleft physical reality\textquotedblright\ than $\mathcal{\hat{D}}%
$. Like a microstate of classical statistical mechanics, each $\left\vert
\varphi_{k}\right\rangle $\ would be associated with a particular realization
of the ensemble;\ it would represent\ an individual system, occurring in the
ensemble with a relative frequency $q_{k}$ (``realist interpretation'', 
for more on this, see e.g. ~\cite{deMuynck}).\ The probabilities that appear
 through the pure states $\left\vert \varphi_{k}\right\rangle
$\ and through the weights $q_{k}$\ would have two different natures, the
former, \textquotedblleft purely quantal\textquotedblright, being a property
of the object, and the latter resulting from our lack of knowledge.

That this is a false idea was stressed by de Muynck~\cite{deMuynck}.
On the one hand, contrary to what happens in classical
statistical mechanics, the decomposition into a given mixture
$\mathcal{\hat{D}}$\ is in general not unique.  For
instance, it is impossible to distinguish whether the unpolarized state of a
spin $\frac{1}{2}$\ describes a population of spins pointing (with the same
weight) in the $\pm z$-directions, or in the $\pm x$-directions, or
isotropically in arbitrary directions.\ Thus no physical meaning can be given to\ pure states
$\left\vert \varphi_{k}\right\rangle $\ that would underlie $\mathcal{\hat{D}%
}$,\ since\ they cannot be defined unambiguously. On the other hand, a pure
state has no more, no less \textquotedblleft physical
reality\textquotedblright\ than a mixture, since
 it is also just a mathematical tool which allows us to predict any expectation
value for a statistical ensemble of systems, and to evaluate any probability
~\cite{vKampen,Balian89}.
Indeed, the non-commutativity of the observables which represent the
physical quantities in quantum mechanics entails an {\it irreducibly
probabilistic} nature of the theory. Within our quantum approach, 
we therefore refrain from imagining
a more ``fundamental'' description which would underlie
the statistical interpretation and would apply to individual systems.
~\footnote{Such a description may exist though, be it not at the quantum level,
but Beyond the Quantum.}

Anyhow, even if one wished to deal only with pure initial states, each one
leading through a unitary transformation to a pure final state,
all conclusions drawn from the form Eq. \ref{Dtf=} of the final density operator
would remain valid in a statistical sense. We are interested only in generic experiments;
very unlikely events will never be observed, due to the huge value of the recurrence times.
The situation is comparable to that of a classical gas:
individual trajectories are reversible, and some of them may exhibit a 
pathological behavior. However,
the consideration of the whole bunch of possible trajectories associated with the
physical situation leads to statistical properties that agree with the 
more feasible theoretical analysis in the
language of statistical mechanics -- here of density operators.

Note finally that the lack of off-diagonal blocks in the expression Eq. \ref{Dtf=} of the
final state S+A allows us to use for this state a classical probabilistic description, 
with classical correlations.
In the first stages of the process, the density operator $\hat{\cal D}(t)$
presents all the singular features of quantum mechanics that arise from the
non-commutativity of the physical quantities.
The dynamics of the large system S+A modifies, as usual in statistical mechanics,
the qualitative properties, letting, for instance, irreversibility emerge from
reversible microscopic equations of motion. Moreover, here, we witness the {\it emergence
of standard probabilistic}, scalar-like correlations between S and A in the final state
from the quantum description in which $\hat{\cal D}(t)$ behaves as an operator-like
probability distribution describing a statistical ensemble.

\section*{Acknowledgement}
The authors are grateful for discussion with Willem de Muynck.

\vfill

\begin{thebibliography}{99}
\bibitem {merv}
For our quadratic interaction the paramagnet is unstable at low temperatures, and
this is a drawback for our purposes, since the state will decay without
coupling with the tested spin. More precisely, for $p=2$ and $g=0$ the
instability time for the paramagnetic state can be estimated to be or order
$\ln N$; if $N$ is so large that this time is larger than the measurement time
$\propto1/\gamma$, then the model with $p=2$ is still reasonable as an apparatus.
\end{thebibliography}
 \end{document}